\documentclass[a4paper,12pt]{article}
\usepackage{cb}
\usepackage{epsf}

\begin{document}

\title{Visualizing Conformations in Molecular Dynamics}
\author{Christoph Best\thanks{\ \ %
           current address: John von Neumann Institute for Computing,
                            54245 J\"ulich, Germany.\protect\\\hspace*{5mm}
                            E-Mail: {\em c.best@computer.org\/}\protect\\
             \hspace*{5mm} Keywords: Molecular Dynamics, Conformations, Visualization, 
             Cluster Analysis. \protect\\ \hspace*{5mm}
             },\,
        Hans-Christian Hege}
\date{{}}
\date{March 1999}
\maketitle
%

\newcommand{\V}[1]{{\bf #1}}
\vspace*{-1cm}
\begin{abstract}
The Monte Carlo simulation of the dynamics of complex molecules
produces trajectories with a large number of different configurations
to sample configuration space.  It is expected that these
configurations can be classified into a small number of conformations
representing essential changes in the shape of the molecule.  We
present a method to visualize these conformations by point sets in the
plane based on a geometrical distance measure between individual
configurations.  It turns out that different conformations appear as
well-separated point sets.  The method is further improved by
performing a cluster analysis of the data set.  The point-cluster
representation is used to control a three-dimensional molecule viewer
application to show individual configurations and conformational
changes. The extraction of essential coordinates and visualization of
molecular shape is discussed.
\end{abstract}

\section{Introduction}

Molecular dynamics simulations on large computers have become
one of the mainstays for investigating the functions of biomolecules.
Using statistical algorithms, they create a large number of snapshots
of the molecule that approximate the expected distribution of
molecular shapes in actual molecular processes. By looking for typical
shapes (conformations) and transition paths in this large data set,
biochemists can learn about the molecular bases of biochemical
processes. Such understanding is important in particular in designing
more efficient medical drugs.

Identifying typical shapes in such a large data set is itself a
difficult task \cite{Huisinga98}. In most simulation, one focuses on a
few characteristic numbers, like the angles or distances between
specific atoms in the molecule, and monitors the change of
these quantities in the simulation. This approach requires some
advance knowledge about which parts of the molecule are important to
the dynamics, and makes it also difficult to perform the analysis
automatically.

To use all the information computed in a molecular dynamics
simulation, one must leave the analysis step again to a computer. We
present here two procedures to aid in this task: a projection method to 
visualize the molecular configurations of a trajectory as a point set
in the plane, and a cluster analysis to identify clusters of similar
configurations in the trajectory. These methods can be applied
automatically to any molecular dynamics trajectory and result in a
tentative identification of conformations in the trajectory.

\section{Procedure}

\subsection{Configurations and feature vectors}

The output of a molecular dynamics simulation is a trajectory,
i.e.~sequence of configurations that depicts the evolution of the
molecule in time.  If the molecule consists of $n$ atoms, the
configuration $x$ is described by $n$ 3-dimensional vectors $\V{x}_i
\in\setR^3$ that specify the cartesian position of each atom in
3-dimensional space. Other information, in particular which atoms are
connected by chemical bonds, is a property of the molecule and usually
does not change during the simulation.

To classify the configurations, we must quantify how much their
geometries differ. We thus assign to each configuration a feature
vector that describes the geometry of the configuration is such a way
that similar configurations have similar feature vectors. However, the
set of $3n$ numbers that make up the cartesian positions of the atoms
is unsuited as identical geometries can appear with different
rotations and translations. While the translational freedom can easily
be fixed by requiring that the center of mass conincides with the origin 
of the coordinate frame, the rotational degree of freedom is extremely
difficult to eliminate. Fixing the axis of inertia can lead to sudden
articifical rotations when the axes become degenerate, while fixing
certain atoms to the coordinate axes always introduces a undesirable
bias. 

A feature vector that is invariant under translations and rotations
and does not introduce any bias can be chosen by considering the set
of intramolecular distances \cite{Cordes95}
\be \label{eq6}
  \{ d_{ij}(x) = |\V{x}_i - \V{x}_j|, \qquad i,j\in 1,\ldots,n \}
  \quad.
\ee
The price to pay is that instead of $3n$ elements, this vector has
now $n(n-1)/2$, but geometrically identical configurations 
have identical feature vectors, and the cartesian distance in
$n(n-1)/2$-dimensional space
is a natural measure of conformational
distance:
\be \label{eq5}
  d(x,y) = \sqrt{\frac{1}{n(n-1)/2}
                  \sum_{i>j} \left( d_{ij}(x) - d_{ij}(y) \right)^2 }
           \quad.
\ee

Another frequent way to choose a feature vector is to use the dihedral
angles between certain atoms as basic degrees of freedom. This is a
natural choice as dihedral angles are the main degrees of motion in
the simulations (atomic distances and bond angles are usually much
more rigid). However, the potential energy that determines the
dynamics of the molecule depends on the spatial distance of the atoms, 
and the relation between dihedral angles and spatial distances is
involved at best. We prefer here to put all information as unbiased as 
possible to the algorithm and depend on it to extract the relevant
degrees of freedom.

The feature vector (\ref{eq6}) is vast compared to the number of
degrees of freedom (in our example molecule, it has 2415 elements as
compared to $70 \times 3 - 6 = 204$ degrees of freedom). Some of its
elements will show little or no fluctuation (e.g.~the ones associated
to the lengths of chemical bonds), others will fluctuate thermally,
and still others will assume different values in different
fluctuations and thus exhibit a double-peaked distribution. To reduce
the thermal noise, we analyze the elements of the feature vector
statistically and select those whose distribution has the largest
width. As thermal fluctuations are smaller than the
conformational changes, this also selects the distances most affected
by conformational changes. A similar procedure has been used by
\cite{Amadei93} to identify essential degrees of freedom in cartesian
coordinate space.

\subsection{Low-dimensional approximations}

The feature vector space is by far too large to be visualized
directly. To capture the major properties of the point set that
represents a trajectory in this space, we seek to visualize it in a
plane, i.e. to assign each configuration a point in the
two-dimensional plane such that the geometrical similarity between
configurations is reproduced as faithfully as possible.  After having
chosen a distance measure (\ref{eq5}) on the trajectory, this reduces
to the general problem of visualizing an arbitrary distance matrix
$D_{ij}$ between a set of $N$ configurations, where $i$ and $j$ now
number configurations.

One choice is to require that the mean quadratic deviation of the
conformational distance from the distance in the plane, given by
\be
  \label{eq2}
  D^2 = \sum_{i>j} \left( |\V{x}_i - \V{x}_j| - D_{ij} \right)
\ee
is minimized by the choice of the points $\V{x}_i$, i.e.~that the derivative of
the quantity with respect to the position of the $k$-th point 
\be
  \label{eq3}
  \frac{\partial D^2}{\partial \V{x}_k} =
  \sum_i \frac{\V{x}_k - \V{x}_i}{|\V{x}_k - \V{x}_i|} \,
  \left( |\V{x}_k - \V{x}_i| - d_{ik} \right)
  = 0
\ee
vanishes.  This equation can be pictured physically by a set of
springs that connect the points and whose natural length is given by
the desired distance between the points.

We solve the minimum problem of (\ref{eq2}) numerically by the
conjugate-gradient method. Though there is no guarantee that the
minimum found by this method is the global one, the minimization
takes place in a $2N$-dimensional space where it is improbable that a
false minimum is stable in all directions. An example of this is the
situation where we have a solution of Eq.~(\ref{eq3}) in $D-1$ dimensions
and then extend the solution space to $D$ dimensions by setting
$x_{i,D} = 0$ for all $i$, which still satifies Eq.~(\ref{eq3}).  However,
this minimum (in $D-1$ dimensions) now turns out to be a saddle point in
$D$ dimensions, where the second derivative of $D^2$ is
\be
  \frac{\partial D^2}{\partial x_{k,D} \partial x_{l,D}}
  = \left\{\begin{array}{cl}
      \frac{d_{kl} - |\V{x}_k - \V{x}_l|}{|\V{x}_k - \V{x}_l|}
      & \mbox{if $k\ne l$} \\
      -\sum_{i\ne k} \frac{d_{ki} - |\V{x}_k - \V{x}_i|}{|\V{x}_k -
        \V{x}_i|}
    \end{array}
    \right.
\ee
In a true minimum, this quantity is positive, thus requiring that
\be
  d_{kl} \ge |\V{x}_k - \V{x}_l| \qquad\mbox{for all $k$, $l$}
\ee
but also
\be
  \sum_{i\ne k} \frac{d_{ki} - |\V{x}_k - \V{x}_i|}{|\V{x}_k -
    \V{x}_i|} \le 0
  \qquad\mbox{for all $k$} \quad.
\ee
This will happen only if the first inequality is an equality, i.e.~if
the solution is complete.

Another widely used low-dimensional approximation is based on the
singular-value decomposition (SVD) of the feature matrix
\cite{Frieze98}. 
Let $a_{ij}$ be
the feature matrix of $i=1,\ldots,n$ objects with $j=1,\ldots,m$
features each. (In our example, $n$ is the number of configurations
while $m$ is the number of intramolecular distances.) The
singular-value decomposition expresses this matrix as a series
\begin{equation}
\label{eq1}
  a_{ij} = \sum_k \lambda_k u^{(k)}_i v^{(k)}_j
\end{equation}
where $\V{u}^{(k)}$ and $\V{v}^{(k)}$ are $n$- and $m$-dimensional,
resp., orthonormalized basis vectors, and $\lambda_k$ gives the
weight of $k$-th term. The number of terms in the series is the
rank of the matrix, it is at most the lower of $n$ and $m$.

The relation between singular-value decomposition and point sets in
low-dimensional space can be seen by calculating the distance between
feature vectors in terms of the SVD:
\begin{eqnarray}
  D_{ij} &=& \sum_{k} (a_{ik} - a_{jk})^2 \nnm\\
  &=& \sum_k \left( \sum_l \lambda_l (u^{(l)}_i - u^{(l)}_j) v^{(l)}_k 
             \right)^2 \nnm\\
  &=& \sum_k \lambda_k^2 \left(u^{(k)}_i - u^{(k)}_j \right)^2
\end{eqnarray}
when orthonormality of $v^{(k)}$ is taken into account. Thus the
vectors
\be
  \left\{ \lambda_k \, u^{(k)}_i : k=1,\ldots,m \right\}
\ee
can be interpreted as specifying the cartesian positions in
$m$-dimensional space of the $i$-th data point. When we chose
$\lambda_k$ in decreasing order, truncating the series after
singular-value decomposition after $l$ terms will lead position
vectors in $l$-dimensional space that are best approximations in a
linear sense.

The major difference between the two approaches is that the SVD
performs the approximation is a linear manner: When the dimension of
the approximation space is decreased from $D$ to $D-1$, the new
approximation is simply obtained by orthogonally projecting out the
last coordinate. In contrast, in the approximation obtained from
minimizing (\ref{eq2}), the nonlinearity introduced by the square root 
redistributes some of the ``lost'' distance in the remaining
dimensions.

\subsection{Cluster analysis}

Cluster analysis \cite{Jain88,Anderberg73} is a statistical method to
partition the point set into disjoint subsets with the property that
the points in a subset are in some sense closer to each other than to
the remaining points. There are several different ways to make this
statement mathematically precise. We choose the notion of minimum
residual similarity between clusters which leads to a natural
formulation of the problem in terms of eigensystem analysis and to a
heuristic algorithm for its solution.  This spectral method goes back
to works by Donath and Hoffmann \cite{Donath72,Donath73} on graph
partitioning in computer logic and Fiedler \cite{Fiedler75,Fiedler75b}
on acyclic graphs and was later picked up by Hendrickson
\cite{Hendrickson95b}. Other cluster analysis methods based on neural
networks or fuzzy clustering have also been applied to molecular
dynamics simulations \cite{Gordon92,Karpen93}. 

Amadei {\em et.al.\/} \cite{Amadei93} went further by introducing the
concept of essential dynamics in which the coordinate space of the
molecule is split into a small essential subspace and a larger
non-essential subspace. They assumed a linear factorization of the
coordinate space and identified essential coordinates by large second
moments of their distribution, assuming that these distributions are
mainly non-Gaussian double-peaked shapes.

To be as flexible as possible, we assume that a similarity measure 
\be 
0 \le a_{ij} \le 1, \qquad 1\le i \le n 
\ee 
is given between the $n$ data points, where $a_{ij}=0$ indicates
complete dissimilarity and $a_{ij}=1$ complete identity of
configurations $i$ and $j$. The residual similarity of a cluster $C
\subset \{1,\ldots,n\}$ characterizes how similar elements of the
cluster are to elements outside the cluster 
\be 
  R(C) = \sum_{i\in C,j\not\in C} A_{ij} \quad.  
\ee
We wish to partition the data set into two subsets
such that this quantity is minimized. Let $a_i$ the characteristic
vector of this partition, with value $a_i=1$ indicating that $i\in C$, 
and otherwise $a_i=-1$. Then the residual similarity can be rewritten
\bea
  \label{eq4}
  R(C) &=& \frac{1}{4} \sum_{ij} (a_i - a_j)^2 A_{ij} \nnm\\
  &=& \frac{1}{2} \sum_{ij} a_i
                            \left( \sum_k A_{ik} \delta_{ij} 
                                   - A_{ij} \right) a_j
  \nnm\\
  &=& (a,Ma)
\eea
with the Laplacian matrix
\be
  M_{ij} = \left\{ \begin{array}{cl}
                   -A_{ij} & \mbox{if $i\ne j$} \\
                   \sum_k A_{ik} & \mbox{if $i=j$}
                   \end{array} \right.
  \quad.
\ee
To find the minimum of the
expectation value $(a,Ma)$ over the vectors
that have element $\pm 1$ only, is a hard combinatorial
problem. However, if we relax the problem and allow real values for
the $a_i$ with the constraint $|a|=1$, the problem is exactly the
problem of finding the
second-lowest eigenvector of the matrix $M$ 
(the lowest eigenvector corresponds to
the solution $a_i \equiv 1$ that does not lead to a proper
partition). Since eigenvectors are orthogonal, the second-lowest
eigenvector satisfies 
\be
  \sum_i a^2_i = 1 \quad\mbox{and}\quad \sum_i a_i = 0 
\ee
and thus guarantees that it will contain both positive and negative
eigenvalues. In graph theory, this eigenvector is called the
characteristic valuation of a graph.

Low-lying eigenvectors of a matrix can be found using iterative 
methods even for moderately large matrices. There is, however, no safe 
way to recover the solution of the combinatorial problem, where $a$ is 
restricted to values of $\pm 1$ from it, but it can be argued that for most
matrices the eigenvector will constitute a good approximation to the
combinatorial problem. We thus map the continuous value $a_i$ to
discrete value $\tilde a_i$ using a threshold $l$:
\be
  \tilde a_i = \left\{ \begin{array}{cl}
                       -1 & \mbox{if $a_i \le l$} \\
                       +1 & \mbox{if $a_i > l$}
                       \end{array} \right. \quad.
\ee
The threshold can now be determined by minimizing the residual similarity
over all possible thresholds. In this way, the minimization problem is 
reduced from $n!$ to just $n$ options, and the characteristic
valuation serves as a heuristic to determine the options that are
taken into consideration.

The measure of residual similarity favors in general
splitting off a single point from the data set since (\ref{eq4})
contains in this case only $n-1$ terms, as compared to $n^2/4$ when
splitting symmetrically. This automatically introduces a quality
control in the splits, as central splits occur only when the cluster
separation is rather favorable, but might also hinder the analysis of
noisy data. However, the special form (\ref{eq4}) was only chosen
to turn the problem into an eigenvalue problem. As the whole
procedure is heuristic in nature, we may well decide to use a
different similarity measure when determining the splitting threshold, 
e.g.~a measure that includes a combinatorial factor
\be
  R(C) = \frac{1}{|C| \, (n-|C|)} \,
         |\sum_{i\in C,j\not\in C} A_{ij} \quad.  
\ee
Which measure is correct depends mainly on the application. The
original measure is stricter in what it returns as a cluster, while
the latter measure favors balanced splittings. In some problems, like
partitioning matrices for processing in a parallel computer, one may
even demand that each split is symmetrical.

Another approach taken frequently in cluster analysis is to use the
singular-value decomposition \cite{Drineas99}. 
If we go back to the feature matrix
$A_{ij}$ and its singular value decomposition (\ref{eq1}), it turns
out that the vectors $u^{(l)}$ correspond to minima of the expectation
value with respect to the feature matrix squared, i.e.  
\be 
  \sum_k \left(\sum_i u_i A_ik\right)^2
  = \sum_{ij} u_i (A_i \cdot A_j) u_j
\ee
where we introduced the row vectors $A_i$ of the matrix $A_{ik}$,
i.e.~the feature vector of data point $i$. Thus 
in this approach the role of the similarity matrix is taken over by
the scalar-product matrix of the feature vectors. The major
differences are\\[-4.4ex]
\begin{enumerate}
\itemsep-0.4ex
\item
The scalar products are not less than or equal to one, but this could
easily be fixed by globally rescaling the scalar-product matrix, which 
does not change the vectors $u^{(l)}$.
\item
The scalar products can be negative. The notion of a scalar-product is 
not of similarity and dissimilarity but rather the trichotomy of
similar, orthogonal, and antagonistic.\\[-4.4ex]
\end{enumerate}
Thus, singular-value decomposition seems suitable for feature vectors
that characterize orthogonal qualities. However, this is not the case
in our feature vectors, so we chose a similarity measure based
upon distance.

After partitioning the data set into two subsets, we proceed to apply
the algorithm again to these subsets. In this way, one obtains a
splitting tree that terminates only when the subset size is smaller than
three. For many applications, such a tree is already quite useful as
it orders the data points in such a way to similar data points are
usually close to each other. 

To identify clusters in the splitting tree, we found that the average
width of the cluster relative to that of its parent cluster gives the
best indication. To calculate the average width of a cluster we use
the Euclidean distance in the high-dimensional space and average over
all distinct pairs of points in a cluster. This quantity relative
to that of the parent cluster basically indicates how much the closer
the points are on average in the subcluster than in the original
cluster and thus how much the split improves the cluster
criterion. Consider e.g.~the situation where there are three
clusters. The first split will result in one correctly identified
cluster and a second pseudo-cluster that encompasses the other two,
but the relative width of the true cluster will be much smaller than
that of the pseudo-cluster. Only after the next split it will be
revealed that the latter consists of two clusters. Typical values for
this quantity are between $0.5$ and $0.8$.

\section{Results}

\begin{figure}[htbp]
\centerline{\epsfxsize=.8\hsize\epsfbox{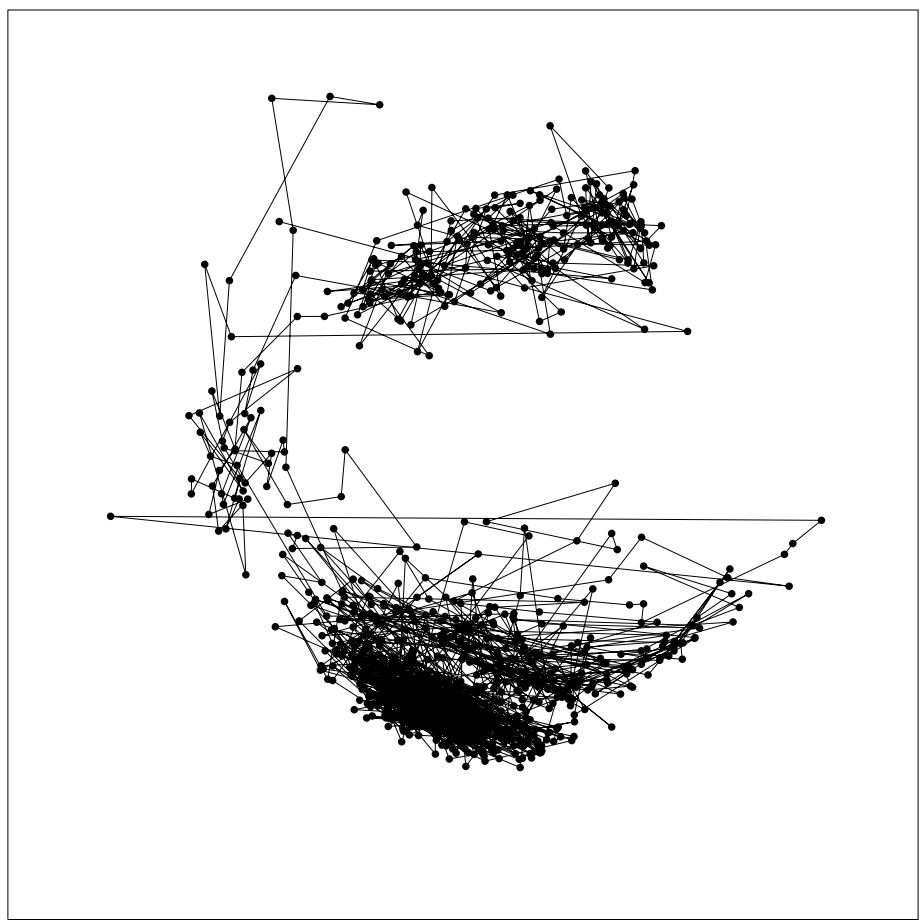}}
\caption{\label{fig1}Two-dimensional map of 1000 configurations chosen from a
molecular dynamics trajectory}
\end{figure}

\begin{figure}[htbp]
\centerline{\epsfxsize=.8\hsize\epsfbox{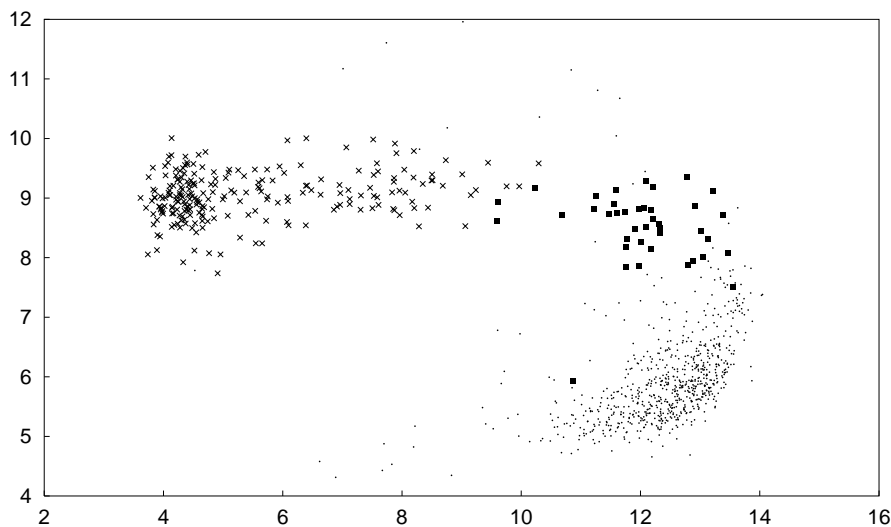}}
\caption{Map of the trajectory in the plane spanned by two typical distances
in the molecule \label{fig2}}
\end{figure}

\begin{figure}[htbp]
\centerline{\hfil\epsfxsize=.4\hsize\epsfbox{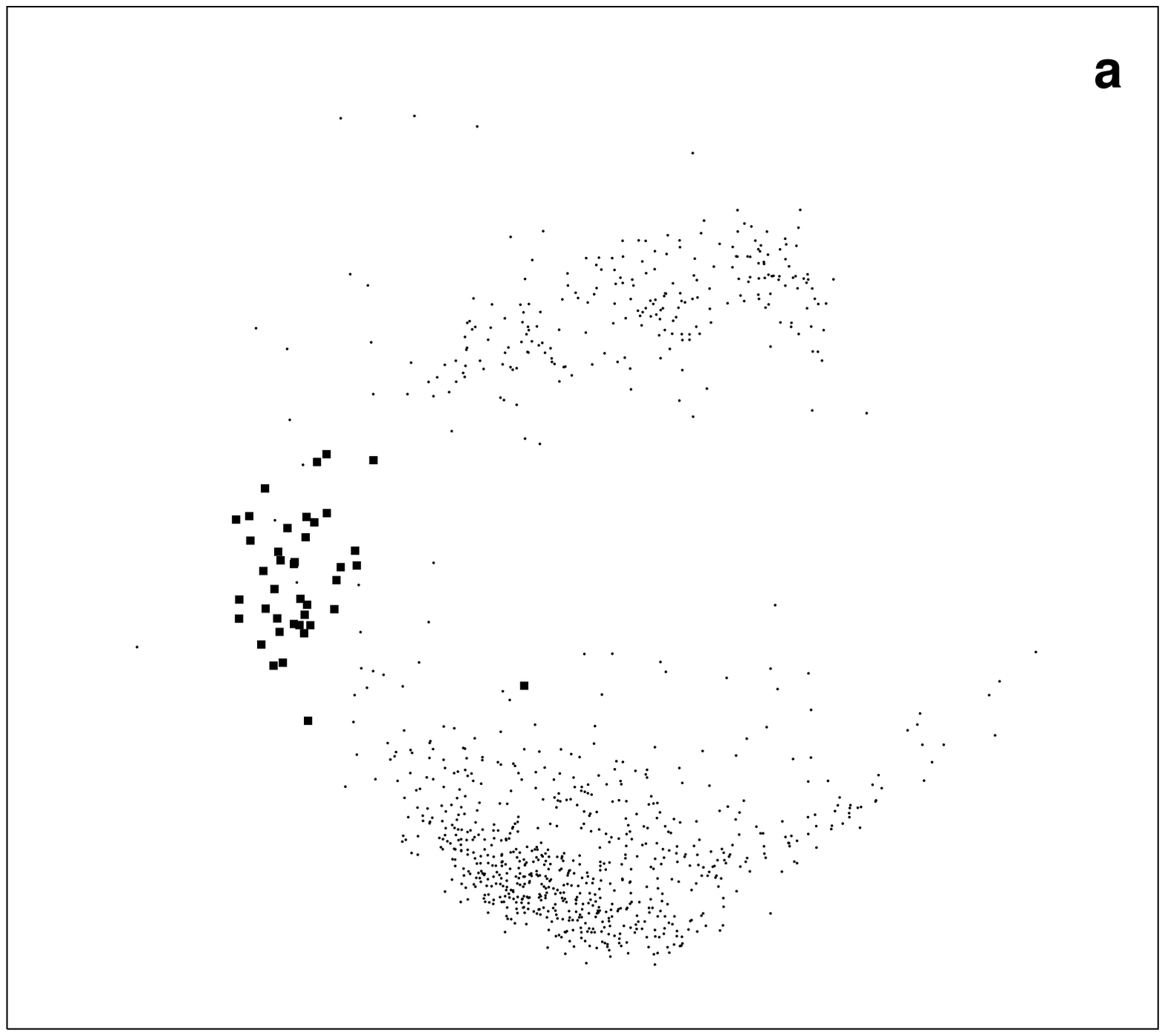}\hfil
\epsfxsize=.4\hsize\epsfbox{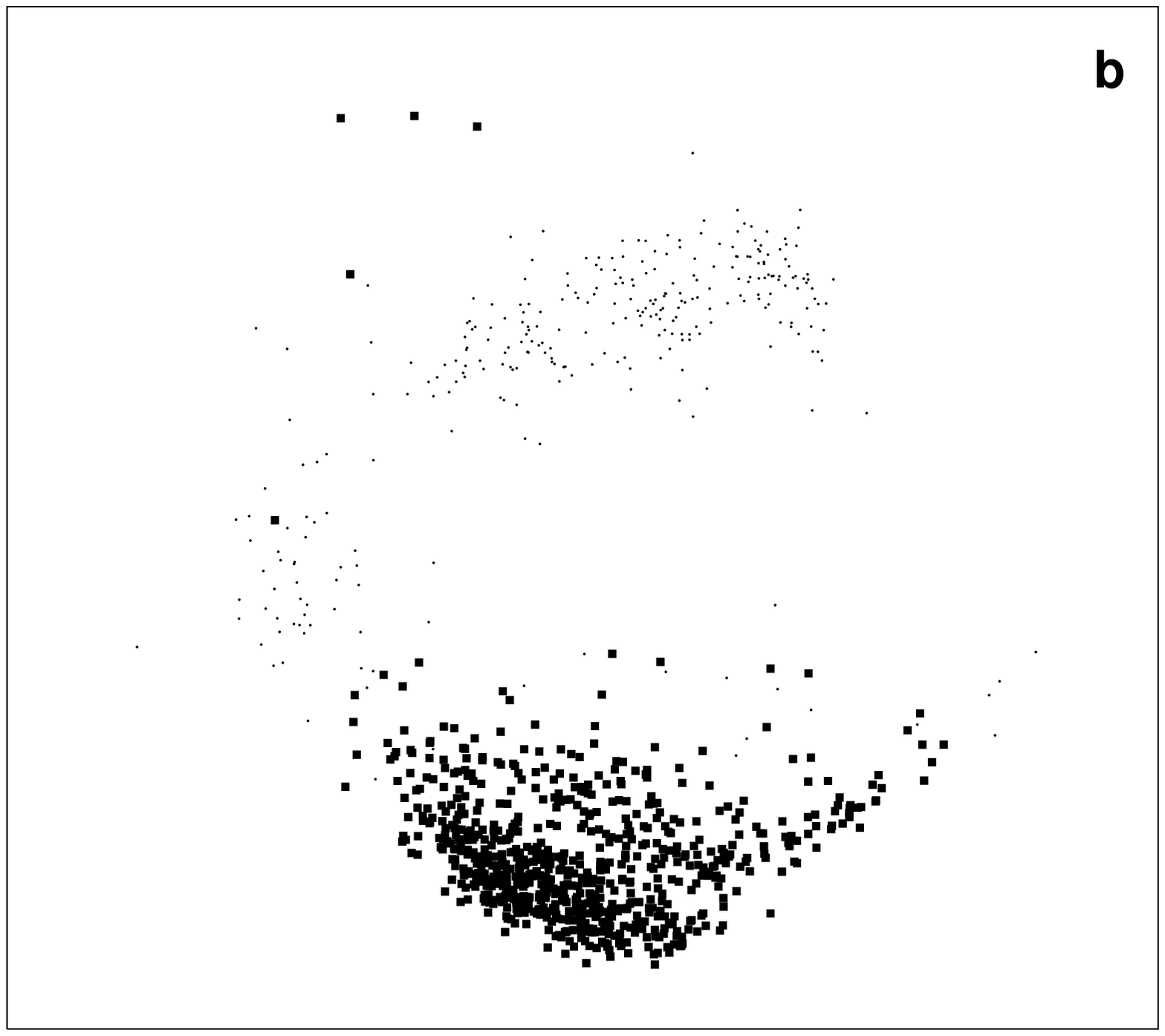}}
\vspace{5mm}
\centerline{\hfil\epsfxsize=.4\hsize\epsfbox{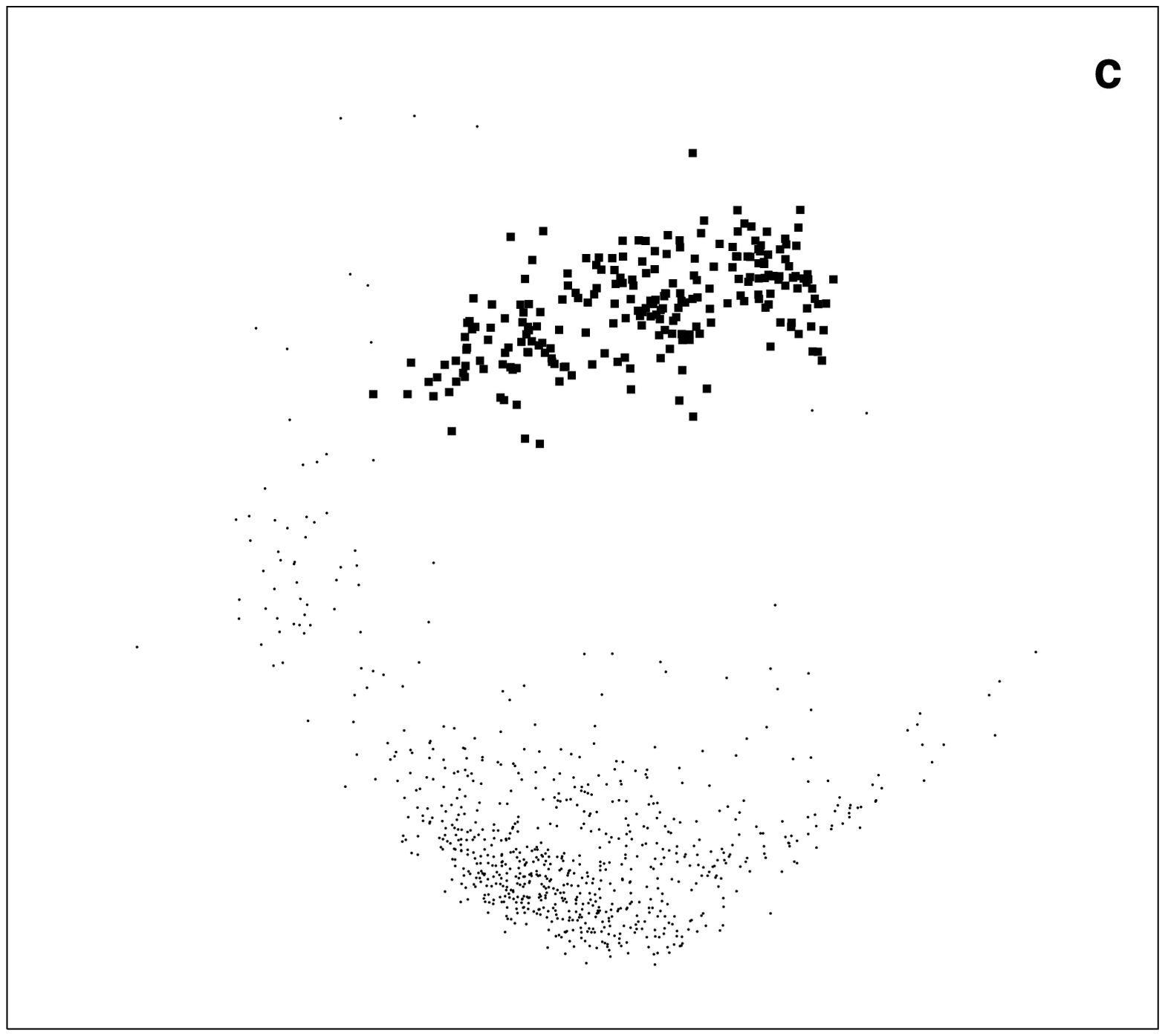}\hfil
\epsfxsize=.4\hsize\epsfbox{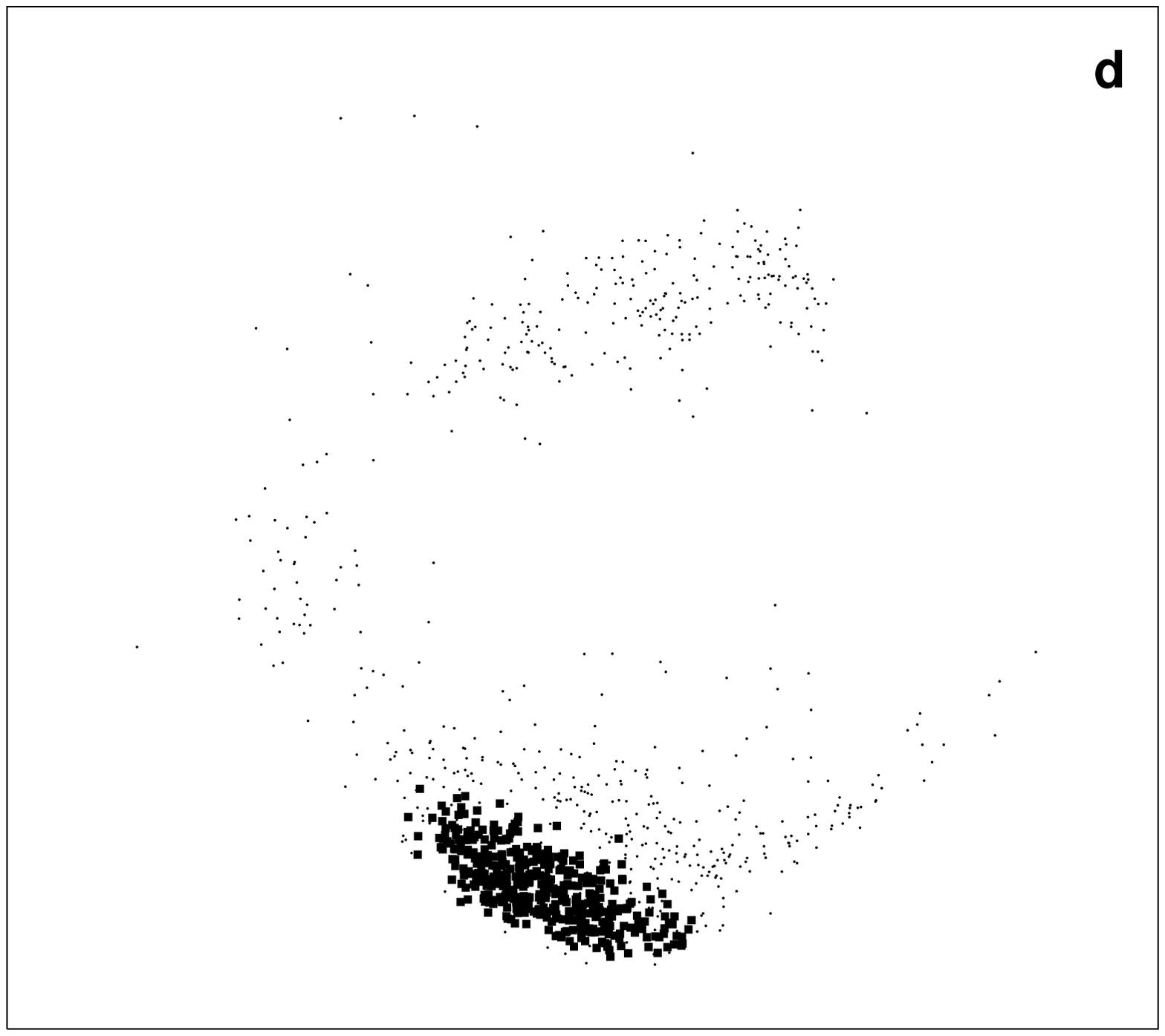}}
\vspace{5mm}
\centerline{\hfil\epsfxsize=.4\hsize\epsfbox{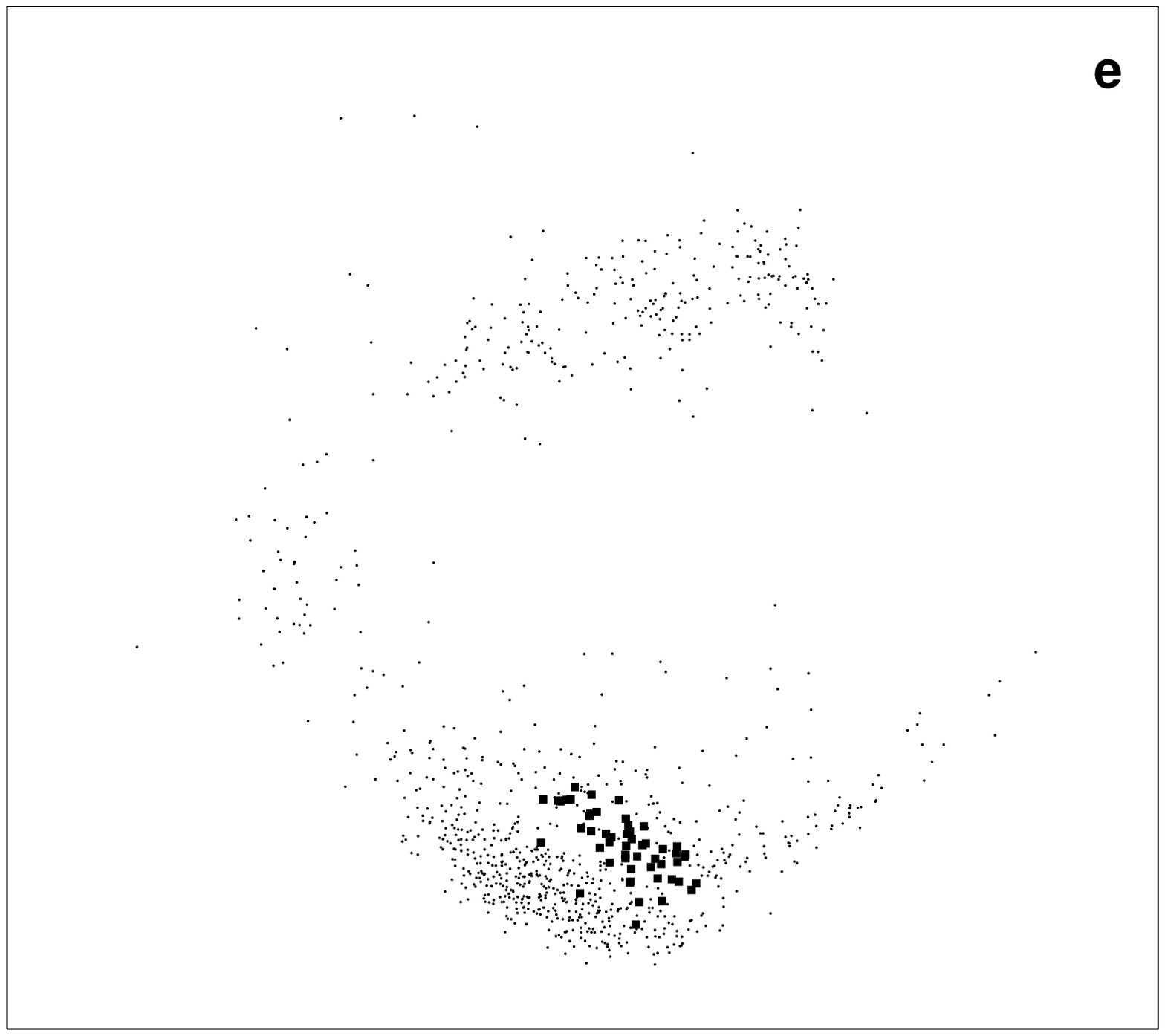}\hfil
\epsfxsize=.4\hsize\epsfbox{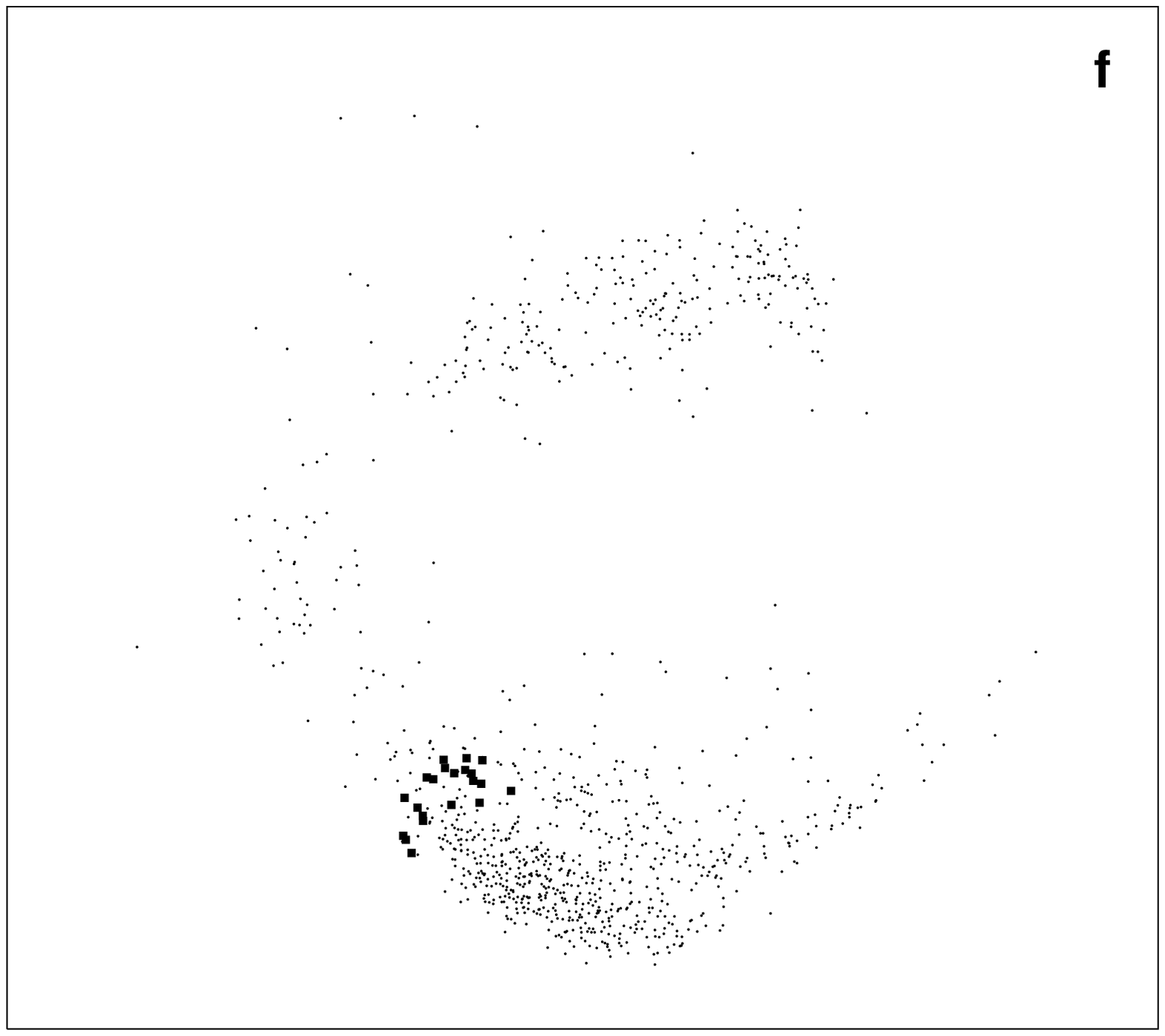}}
\vspace{5mm}
\caption{Different clusters identified in the trajectory by the
  clustering algorithm. \label{fig3} Figures a, b, and c show
  the decomposition of the trajectory into three conformational
  clusters , while
  d, e, and f show the substructure of one such cluster.}
\end{figure}

We apply our methods to a molecular dynamics simulation of the
molecule {\em adenylyl(3'-5')cytidylyl(3'-5')cytidin\/} in vacuum
\cite{Fischer98}.  This is a very simple tri-ribonucleotide, consisting
of three residues and 70 (effective) atoms. The simulation was
performed using the GROMOS96 \cite{GROMOS} extended atom force field.
For the analysis, we chose a subset of 1000 configurations
equidistantly from the trajectory.

Fig.~\ref{fig1} shows the two-dimensional map of the trajectory found
by minimizing (\ref{eq2}). The points are connected by a line in the
same sequence as they are generated in the Monte Carlo simulation.
This information does not enter in determining the locations of the
points in the plane, so the fact that the line segments are rather
short indicates that point adjacent in the trajectory are mapped to
nearby points in the plane and thus are recognized as geometrically
similar by the algorithm.  The one pair of lines that crosses nearly
the whole plane horizontally is actually made up of the first three
data points and therefore a transient effect before the molecule
became equilibrated.

We immediately notice that there are at least three clearly different
groups of points which constitute conformations in a geometrical
sense, i.e.~subsets of the trajectory with similar geometrical
properties.  That they are also dynamical conformations can be seen
from the fact that the connecting line of the points only very rarely
crosses from one point group into the next. This again confirms that
the two-dimensional layout in the plane chosen by the algorithm
represents correctly the dynamics of the system.

The representation of Fig.~\ref{fig1} can be compared to a
representation where the dimensionality of the system is reduced by
chemical understanding. As the system consists of three residues,
most of the conformational dynamics can be assumed to be in the
geometrical layout of the residues. This can be described by only
three numbers, and we chose two of them to create the two-dimensional
representation shown in Fig.~\ref{fig2}. This picture is similar to
Fig.~\ref{fig1} in that there appear approximately three distinct
point groups, and it can be verified that they correspond to the point
groups from Fig.~\ref{fig1}. However, the separation of the point
groups is less clear than in Fig.~\ref{fig1}. This indicates that the
conformational dynamics is not simply the motion of the {\em centers}
of the residues, but there are also smaller rearrangements in the
residues themselves that are correlated to the large-scale motions. By
considering an unbiased measure for geometrical similarity, all those
little rearrangements enter and reinforce the distance between
conformations in the plane.

\subsection{Cluster analysis}

Applying the cluster algorithm to the similarity matrix of the
trajectory, the first few splits remove 22 isolated points before the
small cluster shown in Fig.~\ref{fig3}a with 40 points and a relative
width of 0.46 (both compared to its immediate predecessor and to the
initial point set) shows up. The remaining points are split some steps
further into a cluster with 698 points shown in Fig.~\ref{fig3}b and
another cluster with 230 points shown in Fig.~\ref{fig3}c with
relative widths of about 0.53. After some more steps, the larger
subcluster is broken into three subclusters with 388, 52, and 21
points, resp., and relative widths of 0.91, 0.73, and 0.61, resp., as
shown in Fig.~\ref{fig3}d, e, and f. Similarly, the smaller subcluster 
also separates into three weak subclusters.

The splitting line of the large cluster at the bottom is also visible
in Fig.~\ref{fig1}. Such a pattern usually indicates that beside a
large conformational change that induces the three clearly visible
clusters, where the middle cluster is clearly a transitional state,
there is another smaller conformational change, possibly in one of the 
glucose rings, independent of the larger one. As it only affects a
small part of the molecule, the conformational distance is smaller and 
is then imprinted like a fine structure on the clusters. That such
changes are visible in the plot is an advantage from considering all
atom coordinates without bias.

\section{Conclusions and Outlook}

We have demonstrated a method for projecting a molecular dynamics
trajectory onto a plane to capture the conformational structure of the
trajectory. Conformations can in this way be easily identified by
visual inspection. Cluster analysis on the full conformational
distance matrix also revealed these clusters, but also allowed to
discern fine structure inside the clusters caused by smaller
conformational changes.

To simplify the analysis of a trajectory, we have created a Java
application that reads the output files of the combined plane
mapping/cluster analysis program and displays the two-dimensional map.
This program interacts directly with an Open Inventor molecular
visualization application by means of a Unix pipe. Whenever the user
selects a point in the plane, the corresponding configuration is shown
in the visualization program. The user can also choose to display
identified clusters using different colors in the map.

Identifying which parts of the molecule are responsible for different
conformations is a much more difficult problem. We use a visualization
application that allows the user to form groups of atoms that are
visualized by ellipsoids. In this way, a molecule can be easily
reduced to its functional groups where it is much easier to spot
conformational changes. However, small conformational changes as those 
that show up as a fine structure on the plane map are easily lost in
this representation.

\begin{figure}[htbp]
\centerline{\epsfxsize=.8\hsize\epsfbox{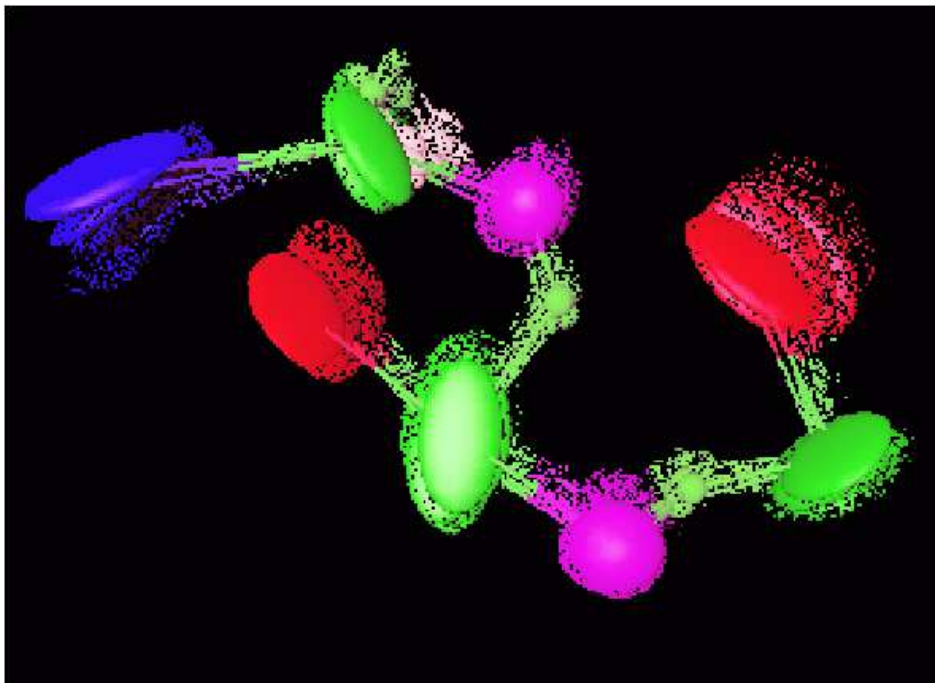}}
\caption{Representation of the collective motion by an OpenGL fading
effect \label{fig4}}
\end{figure}

As a first attempt at aiding the eye in discovering unusual motions of
the molecule, we implemented a simple OpenGL effect in the
visualization application that allows to blend several frames of an
animation in the hope that large changes stand out more clearly in
this representation.  Fig.~\ref{fig4} shows one such
picture. Certainly more research can be expended on how to identify
and visualize the essential degrees of freedoms.

The concept of essential molecular dynamics has been introduced to
reduce the number of degrees of freedom in the simulation. Both the
plane map and the cluster analysis can be used to inflict new
coordinates upon the system. For the point map, these are simply the
$x$ and $y$ positions of the configuration in the plane. Once a
certain point map has been established, new configurations can be
fitted into the plane by minimizing the residual distance while
keeping all other points fixed. Similarly, the cluster analysis
assigns to each configuration a position in the tree that can be seen
as a (discrete) essential coordinate. How such essential coordinates
can be reintroduced into the dynamics of the system is still an open
question.

\end{document}